\documentclass[12pt]{article}
\usepackage{epsfig}
\usepackage{graphics}
\begin{document}

\title{The Perfectly Matched Layer for nonlinear and matter waves}
\author{C. Farrell, U. Leonhardt\\
School of Physics and Astronomy, University of St Andrews,\\
North Haugh, St Andrews, Fife, KY16 9SS, Scotland}
\maketitle

\begin{abstract}
We discuss how the Perfectly Matched Layer (PML) can
be adapted to numerical simulations of 
nonlinear and matter wave systems, such as
Bose-Einstein condensates. We also present some examples 
which illustrate the benefits of using the PML in the
simulation of nonlinear and matter waves.
\end{abstract}

\newpage

\section{Introduction}
The Perfectly Matched Layer, or PML, is an absorbing boundary
condition (ABC) which was introduced in a paper by Berenger
\cite{berenger}. ABCs are necessary when solving partial
differential equations (PDEs) using finite-difference time-domain
(FDTD) methods or finite element methods (FEM) over an open
region. An efficient ABC will save computer memory and processing
time by truncating the domain and will minimize numerical
reflections from the domain truncation. The PML is reflectionless
in theory, though some small reflection does occur in practice due
to discretization. The magnitude of this reflection is
considerably smaller than that of other ABCs, however (for one
comparison, see Ref. \cite{andrew}). 
It has been extensively used in
the field of electromagnetics 
(see Refs. \cite{mittra,demoerloose,wu,veihl,xu} for
example) and has also found application in acoustical and
geophysical work 
\cite{hastings,liu,zeng1,zeng2,oguz}. 
To our knowledge, a PML has never been applied in
numerical simulations of 
matter waves, such as Bose-Einstein condensates.

Imagine that we wish to simulate fluid flowing through a tube with
a constriction and becoming supersonic, such as happens in a
rocket engine or a Laval nozzle, see Fig. 1.
\begin{figure}[htbp]
\begin{center}
\centerline{\epsfig{file=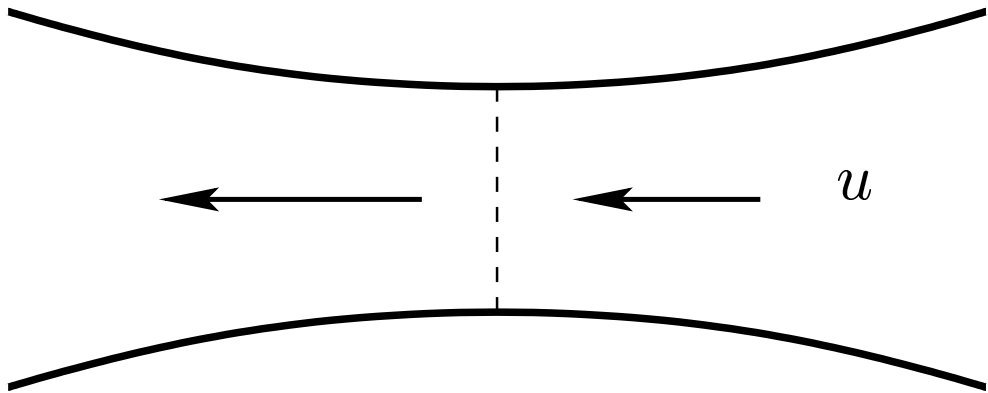}}
\end{center}
\vspace{0ex} \caption{Schematic illustration of a supersonic nozzle.
The flow velocity $u$ exceeds the speed of sound at the throat of the nozzle}
\end{figure}
Such a system could be used as a laboratory analogue of a black
hole \cite{unruh, abh}, using a Bose-Einstein condensate for the
fluid and laser light to provide the constriction. Acoustic waves
which have passed the constriction cannot propagate back fast
enough to return through it because of the fluid's supersonic
flow. We can say that a sonic horizon, which is analogous to the
event horizon of a black hole and should share many of its
properties, such as the production of Hawking-like radiation, is
formed at the narrowest point in the tube. A supersonic nozzle is
difficult to model computationally, because if we would
impose the usual periodic boundary conditions we would need
to slow down the flow to subsonic speed.
The process of turning a
supersonic flow to a subsonic one is even more unstable than the
converse \cite{LKO}
and so periodic boundary conditions are unsuitable. This
is where the PML technique becomes useful. Placing a perfectly
'black' layer at the supersonic end of the system allows us to
model the condensate flow with much greater physical accuracy, by
absorbing the supersonic matter waves and so simulating their
propagation into free space.

Another situation where the PML proves useful is in the simulation
of the generation and propagation of waves with attractive
nonlinear interaction, such as solitons. The use of a PML layer at
the edge of the computational domain can absorb the solitons,
allowing us to ignore their fate and to focus on the area of
interest. The PML technique could be applied to the modelling of
soliton propagation in a BEC \cite{hulet} or in glass fibres
\cite{agrawal}, for example.

\section{Theory and implementation}

\subsection{Theory}
The concept of the PML for nonlinear and matter waves is similar
to that for electromagnetic waves. The essential idea is the use
of transformations which map propagating solutions onto
exponentially decaying evanescent waves in complex space. Thus,
waves travelling in the PML change from propagation in real space
to propagation in imaginary space. The idea of using complex space
for waves originated with Deschamps \cite{deschamps} and the PML
was first interpreted as a complex coordinate stretching by Chew
and Weedon \cite{chewweedon}. Here we show how these ideas can be
applied to nonlinear and matter waves. Consider first the standard
linear Schr{\"o}dinger equation in one dimension,
\begin{equation}
i\frac{\partial \psi}{\partial t} = -\frac{1}{2m} \frac{\partial^2
\psi}{\partial x^2},
\end{equation}
which can be written as
\begin{equation}
i\frac{\partial \psi}{\partial t} = -\frac{1}{2n}
\frac{\partial}{\partial x}\frac{1}{n}\frac{\partial
\psi}{\partial x},
\end{equation}
where $m$, the mass, has been split into two spatially dependent
functions $n$. Suppose that $n$ changes value from 1 at $x=\infty$
to i at $x=-\infty$. Equation (2) has the general solution
\begin{equation}
\psi = \int_0^{\infty}A(\omega) \exp \left(\pm i\int k dx -
i\omega t\right)d\omega, \; k = \pm n \sqrt{2\omega},
\end{equation}
where the term inside the exponential is positive for waves moving
to the left and negative for waves moving to the right. We can
choose $n$ to be, for example,
\begin{equation}
n = \exp \left[\pm i \frac{\pi}{4} \left(1 -
\tanh\frac{x-x_0}{a}\right)\right],
\end{equation}
where the exponential is positive or negative depending on the
direction of propagation of the waves in question. $x_0$ is the
position where the PML starts and $a$ is a parameter which
determines the sharpness of the transition between 1 and $i$,
which should be reasonably gradual, though our simulations
indicate that the transition may be quite steep without any
serious effects. This implies that a matter wave may be stopped
relatively rapidly. Our introduction of $n$ is equivalent to
making the transformation
\begin{equation}
\xi = \int n \mathrm{d}x
\end{equation}
and writing
\begin{equation}
i\frac{\partial \psi}{\partial t} =
-\frac{1}{2}\frac{\partial^2}{\partial\xi^2}\psi.
\end{equation}
Essentially, we have transformed to a coordinate system that
causes waves to exponentially decay as they approach $-\infty$.

The nonlinear Schr{\"o}dinger equation, also known as the
Gross-Pitaevskii equation when used for matter waves, is written,
after scaling through proper coefficient choices,
\begin{equation}
i\frac{\partial \psi}{\partial t} = -\frac{1}{2n}
\frac{\partial}{\partial x}\frac{1}{n}\frac{\partial
\psi}{\partial x} + \epsilon |\psi|^2 \psi, \; \epsilon = \pm 1.
\end{equation}
If $\epsilon$ is positive, the equation represents repulsive
interactions, such as in a dark soliton or BEC; if negative, the
equation represents attractive interactions, as in a bright
soliton. The PML can be extended to deal with both these cases and
also with matter waves provided the incoming wavefront is close to
zero. If this condition is met, the PML acts to keep the incoming
wave small and so the system is approximately in the linear
regime. We perform a coordinate transformation in the same fashion
as Eq. (5) and so we can write
\begin{equation}
i\frac{\partial \psi}{\partial t} =
-\frac{1}{2}\frac{\partial^2}{\partial\xi^2}\psi + \epsilon
|\psi|^2 \psi.
\end{equation}

\subsection{Implementation}
We used a FDTD approach, discretizing via the Crank-Nicolson
method \cite{recipes}. This
method has the advantage of being unconditionally stable due to
its implicit nature and is second-order. The system to be solved
becomes
\begin{eqnarray*}
i\frac{\psi(x,t+\Delta t) - \psi(x,t)}{\Delta t} \!\!\!&=&\!\!\!
-\frac{1}{2}\frac{\nu(x)}{2\Delta
x^2}\left[\nu(x+\frac{1}{2})\left(\psi(x + \Delta x, t + \Delta
t) \right.\right. \\
\!\!\!&-&\!\!\! \left.\psi(x, t+\Delta t)\right)
-\nu(x-\frac{1}{2})\left(\psi(x, t + \Delta t)\right. \\
\!\!\!&-&\!\!\! \left.\psi(x-\Delta x, t+\Delta t)\right)
+\nu(x+\frac{1}{2})\left(\psi(x + \Delta x,
t) \right. \\
\!\!\!&-&\!\!\! \left.\left.\psi(x, t)\right)
-\nu(x-\frac{1}{2})\left(\psi(x, t) - \psi(x-\Delta x, t)\right)\right] \\
\!\!\!&+&\!\!\! \frac{|\psi|^2}{2}\left(\psi(x,t+\Delta t) + \psi(x,t)\right),\\
\end{eqnarray*}
where $\nu(x)=1/n(x)$. The tridiagonal system so produced is
solved using the tridiagonal matrix algorithm or Thomas algorithm
\cite{thomas}. The drawback to this discretization scheme is that
it requires an additional level of detail, at half-steps in space.
This implies the need for a greater number of array elements and
hence more computer operations and memory storage. More
complicated simulations, such as a condensate being forced through
a nozzle by a piston, can take over 12 hours to complete on a
Pentium 4 3Ghz PC.

\section{Numerical Simulations}

\subsection{Attractive interaction: Travelling soliton}

In a 1D computational domain 100 space units wide, travelling
solitons are generated by a boundary condition on the right-hand
side. The virtual start point and velocity of the soliton can be
specified by the user. As can be seen from the simulation data,
the soliton travels until it reaches the PML boundary at $x=10$
where it gets 'stuck', begins to be absorbed and starts to spread
out. The peak value of $\rho = |\psi|^2$ quickly falls to less
than $0.03\%$ of the initial value.
\begin{figure}[htbp]
\begin{center}
\centerline{\epsfig{file=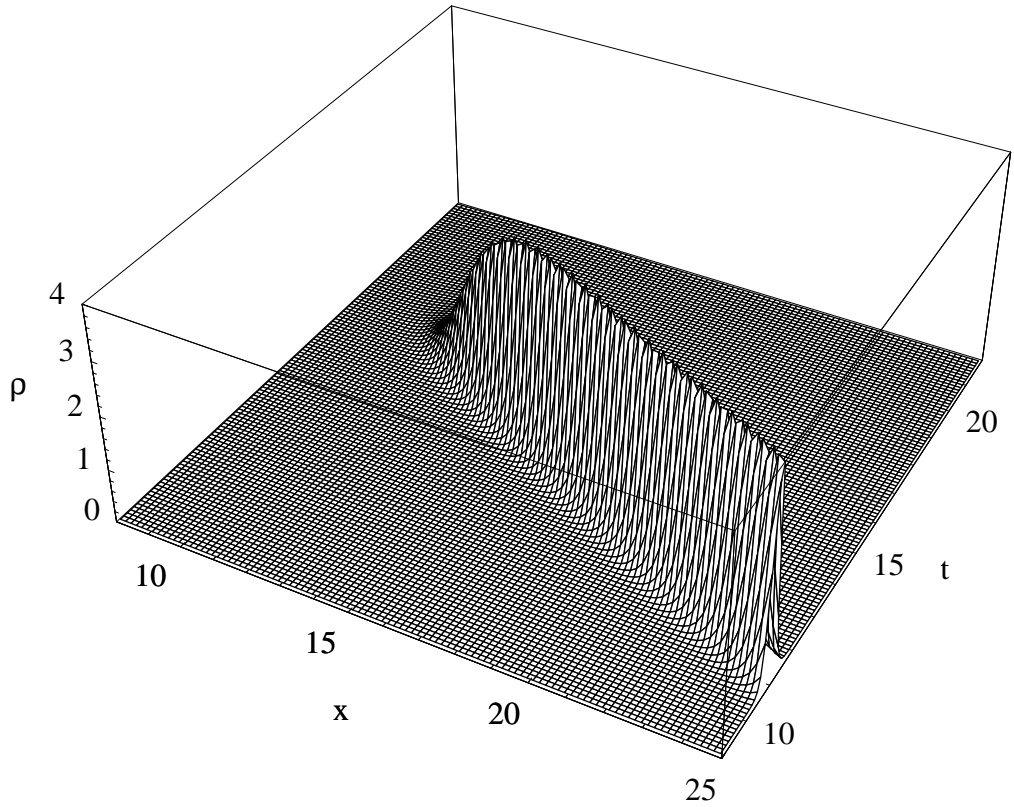}}
\end{center}
\vspace{0ex} \caption{The soliton propagates to the left until it
reaches the PML at $x=10$. Here it becomes 'stuck' and $\rho$
drops sharply, to about 0.03\% of the original value.}
\end{figure}

\subsection{Repulsive interaction: Free expansion of a BEC}
The simulation results show a condensate expanding without
instabilities in a domain 100 space units in size, even with a
relatively small PML (5 space units on the left- and right-hand
sides). Without the PML, monotonically growing instabilities occur
in the expanding condensate and threaten the stability of the
program. When we compared the expansion with the PML in place to
expansion in a computational domain large enough that the
expanding condensate does not touch its sides, no appreciable
difference was evident, which indicates that the PML is
functioning correctly.
\begin{figure}[htbp]
\begin{center}
\centerline{\epsfig{file=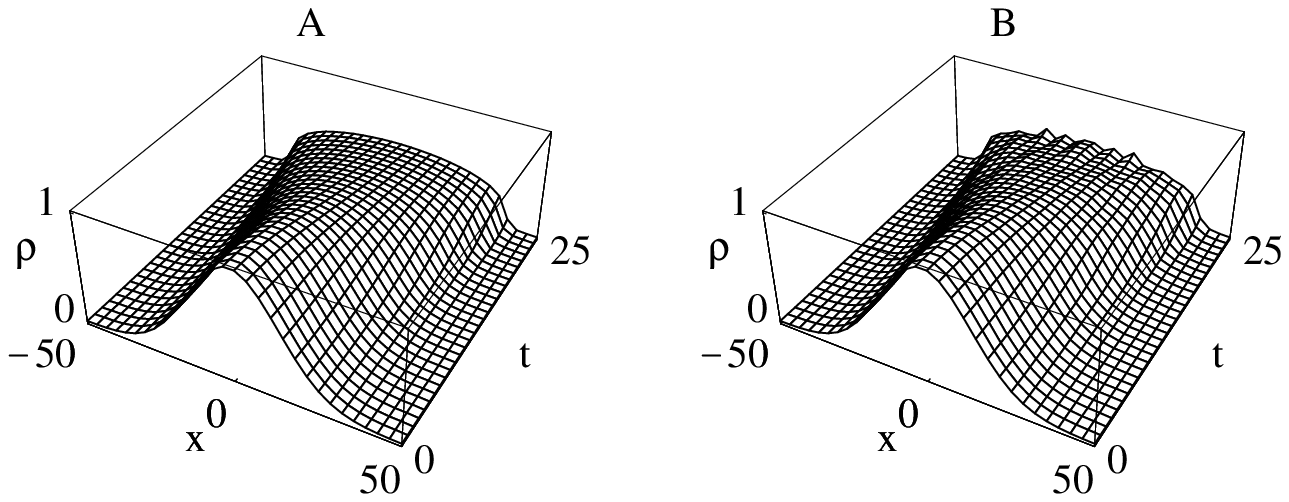}}
\end{center}
\vspace{0ex} \caption{These pictures show the expansion of a
condensate in a domain 100 space units wide. In (A) the PML is
placed in a 5-unit-wide region on the left- and right-hand sides
of the domain, while in (B) the PML is not present, which causes
monotonically growing instabilities to be seen as time proceeds.
The expansion with the PML is practically identical to the
expansion in a 1000-unit-wide domain, where the condensate does
not touch the sides of the computational domain.}
\end{figure}

\section{Conclusion}
As our numerical simulations have shown, the Perfectly Matched
Layer can be used to good effect in nonlinear and matter wave
simulations by truncating the domain, both for repulsive and
attractive interactions. The PML can be put to a wide variety of
uses, for example in the modelling of sonic analogues of a black
hole, expanding BECs or soliton pulses propagating in free space
and fibres. One interesting feature is that the transition between
real $n$ and imaginary $n$ may be relatively sharp in practice,
which indicates that even supersonic condensates may be stopped
quite abruptly. Finally, the PML is of course capable of being
extended to problems in more than one dimension. Here the variable
$x$ in Eqs. (1)
and (2) should be changed to refer to the normal direction of the
boundary, while all other coordinates are unchanged.

\end{document}